\documentclass[twocolumn,superscriptaddress,english,prl]{revtex4-1}
\usepackage[colorlinks=true,urlcolor=blue,citecolor=magenta,linkcolor=blue]{hyperref} 

\usepackage{graphicx}
\usepackage{txfonts}

\def\eqa{\begin{eqnarray}}
\def\eea{\end{eqnarray}}
\newcommand{\eq}{\begin{equation}}
\newcommand{\ee}{\end{equation}}

\newcommand{\Eq}[1]{Eq.~(\ref{#1})}

\begin{document}

\title{Recommender Engine for Continuous Time Quantum Monte Carlo Methods}
%\title{Recommender Engine, Molecular Gases, and Continuous-time Quantum Monte Carlo}

\author{Li Huang}
\affiliation{Science and Technology on Surface Physics and Chemistry Laboratory, P.O. Box 9-35, Jiangyou 621908, China}
\author{Yi-feng Yang}
\email{yifeng@iphy.ac.cn}
\affiliation{Beijing National Lab for Condensed Matter Physics and Institute
of Physics, Chinese Academy of Sciences, Beijing 100190, China}
\affiliation{Collaborative Innovation Center of Quantum Matter, Beijing 100190, China}
\affiliation{School of Physical Sciences, University of Chinese Academy of Sciences, Beijing 100190, China}
\author{Lei Wang}
\email{wanglei@iphy.ac.cn}
\affiliation{Beijing National Lab for Condensed Matter Physics and Institute
of Physics, Chinese Academy of Sciences, Beijing 100190, China}

\begin{abstract}
Recommender systems play an essential role in the modern business world. They recommend favorable items like books, movies, and search queries to users based on their past preferences. Applying similar ideas and techniques to Monte Carlo simulations of physical systems boosts their efficiency without sacrificing accuracy. Exploiting the quantum to classical mapping inherent in the continuous-time quantum Monte Carlo methods, we construct a classical molecular gas model to reproduce the quantum distributions. We then utilize powerful molecular simulation techniques to propose efficient quantum Monte Carlo updates. The recommender engine approach provides a general way to speed up the quantum impurity solvers.
%By fitting the configurations to the energy function of a classical motnotonic gas model, we find an efficient approximate to the distribution function of the quantum problem. 
%to increase the acceptance probability. We present a way to speed up the 
%Generalizations of the proposed approach to lattice models or in the dynamical mean-field theory calculations are straightforward. 
\end{abstract}
\maketitle

%QMC in general 
At the heart of every quantum Monte Carlo (QMC) method is a quantum to classical mapping. One has to find a classical representation of the quantum system to program it into the classical computers~\cite{gubernatis2016quantum}. Since the mapping is not unique, there are various QMC methods suitable for bosons, quantum spins or fermionic quantum many-body systems~\cite{Blankenbecler:1981vj, Sandvik:1991tn, PhysRevLett.77.5130, Prokofev:1998tc, 1999PhRvB..5914157S, Evertz:2003ch, Kawashima:2004clb, Rubtsov:2005iw, Werner:2006ko, Gull:2011jd, 1999PhRvL..82.4155R, Iazzi:2015hi, 2015PhRvB..91w5151W}. The classical representations behind these QMC methods can be classical spins, particles or polymers, etc. In this unified quantum to classical mapping point of view, various QMC methods differ by the implementation details, but all share the same principle. 

%Challenge in QMC
Besides finding the suitable classical representations, another key ingredient of the QMC algorithms is to design efficient strategies to sample the configurations. Successful algorithms~\cite{Prokofev:1998tc, 1999PhRvB..5914157S, Evertz:2003ch} typically identify the collective modes of the effective statistical mechanics problem and make proposals accordingly. However, it is not always easy to construct such efficient updates for generic quantum many-body systems. For example, most of the QMC simulations of the fermionic systems in condensed matter physics still use simple local updates~\cite{Blankenbecler:1981vj, Gull:2011jd}. These updates can be inefficient due to high rejection rate and long autocorrelation times in the Monte Carlo configuration. 

The idea of ``recommender system'' points to a general route to accelerate the quantum Monte Carlo simulations. The recommender system is a broad and active research field~\cite{aggarwal2016recommender} in machine learning. One can build a probabilistic model based on the user's past behavior and suggest favorable products back with high acceptance rates. Similarly, one can model the probability distribution of the QMC configurations with machine learning techniques and propose new efficient Monte Carlo updates accordingly. This core idea has been presented in Refs.~\cite{Huang:2016tf, 1610.03137, 1611.09364}. Models of the recommender systems such as the restricted Boltzmann machine~\cite{Smolensky:1986va, Hinton:2002ic} and the classical spin systems do speed up the Monte Carlo sampling without introducing any bias to the physical results. 

It is, however, not obvious how to extend these ideas to a broader class of modern QMC methods~\cite{Sandvik:1991tn, PhysRevLett.77.5130, Prokofev:1998tc,1999PhRvB..5914157S,  Evertz:2003ch, Kawashima:2004clb, Rubtsov:2005iw,Werner:2006ko, Gull:2011jd, 1999PhRvL..82.4155R, Iazzi:2015hi, 2015PhRvB..91w5151W} in which the number of random variables can fluctuate in the simulation. This paper presents several new ingredients to achieve speedup in these, and in particular, the powerful continuous-time quantum Monte Carlo (CT-QMC) methods. The CT-QMC methods have revolutionized the study of quantum impurity models~\cite{Gull:2011jd} since their invention a decade ago~\cite{Rubtsov:2005iw,Werner:2006ko}. Latest developments~\cite{Iazzi:2015hi, 2015PhRvB..91w5151W} built on Ref.~\cite{1999PhRvL..82.4155R} further extend these successes to lattice fermions~\cite{Wang:2015je, Liu:2015kx}. There were dedicated efforts to optimize the sampling of the CT-QMC methods~\cite{Bourovski:2004jf, Burovski:2006hv, Shinaoka:2014dv, Shinaoka:2015gr, Liu:2015kx} because of their broad impacts. Our innovations include reverse engineer a classical molecular gas model from the CT-QMC configurations and leveraging mature molecular simulation techniques to propose efficient updates back to the CT-QMC simulation. This approach provides a systematic and principled approach to improve the efficiency of the CT-QMC methods. This progress has an immediate impact on the realistic simulation of correlated materials~\cite{Kotliar:2006fl}.  

\begin{figure}[t]
\includegraphics[width=8.8cm]{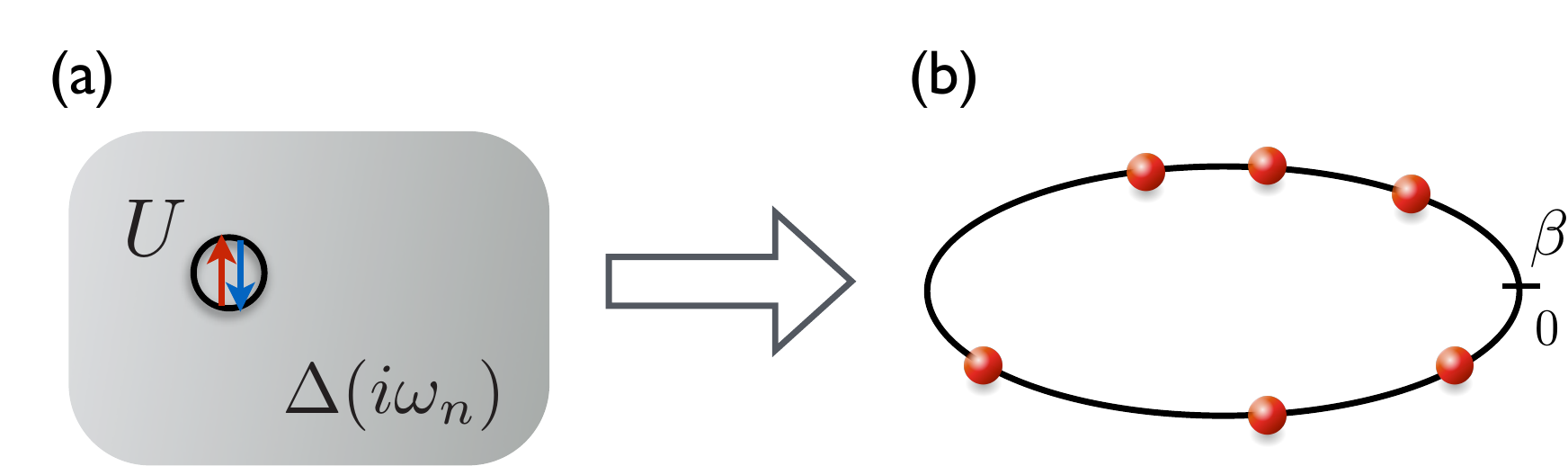}
\caption{(Color online) (a) The quantum impurity problem consists of an impurity with local interaction embedded in a bath of non-interacting fermions. (b) The CT-QMC method maps the quantum impurity model to a one-dimension classical molecular  gas model. Each red dot represents an interaction vertex in the interaction expansion~\Eq{eq:expansion}. The length of periodic imaginary time is the inverse temperature $\beta$. Configurational bias Monte Carlo simulation of the molecular gas recommends efficient updates to the  CT-QMC simulation.   
\label{fig:mapping}}
\end{figure}
  
To make our discussion concrete, we consider the single impurity Anderson model. The action reads~\cite{Anonymous:z_AfEOwS}
\begin{eqnarray}
S & = & -\int_{0}^{\beta} d\tau \int_{0}^{\beta}  d\tau'  \sum_{\sigma=\{\uparrow, \downarrow\}}c^{\dagger}_{\sigma}(\tau) \mathcal{G}^{-1}_{0}(\tau-\tau') c_{\sigma}(\tau') \nonumber \\ & &+ U  \int_{0}^{\beta} d\tau \left(n_{\uparrow}-\frac{1}{2}\right)\left( n_{\downarrow}-\frac{1}{2}\right). 
\label{eq:siam} 
\end{eqnarray}
The model describes a quantum impurity embedded in a non-interacting environment illustrated in Fig.~\ref{fig:mapping}(a). $\beta$ is the inverse temperature and $U$ is the onsite interaction strength. $\mathcal{G}_{0}$ is the non-interacting Green's function of the impurity. In the Matsubara frequency, it reads $\mathcal{G}^{-1}_{0}(i\omega_{n}) = i\omega_{n}+\varepsilon -\lambda^{2}\Delta(i\omega_{n})$, where $\varepsilon$ is the local chemical potential of the impurity and $\lambda$ is the hybridization strength between the impurity and the non-interacting bath. In the following, we consider a bath with a semicircular density of states $\Delta(i\omega_{n}) = {2}/{\left(i\omega_{n}+\sqrt{(i\omega_{n})^{2}-D^{2}}\right)}$~\cite{Anonymous:z_AfEOwS} and set the half bandwidth $D=2$ as the energy unit. Physically, the single impurity Anderson model (\ref{eq:siam}) is relevant to cases of magnetic atoms hosted in a metal or quantum dots coupled to the leads. The model captures rich physical phenomena including local moment forming, Coulomb blockade and the Kondo effect~\cite{KrishnaMurthy:1980wb}. Moreover, solving the quantum impurity model is the computational engine in the dynamical mean-field theory studies of correlated lattice models~\cite{Anonymous:z_AfEOwS}. 

\begin{figure}[t]
\includegraphics[width=\columnwidth]{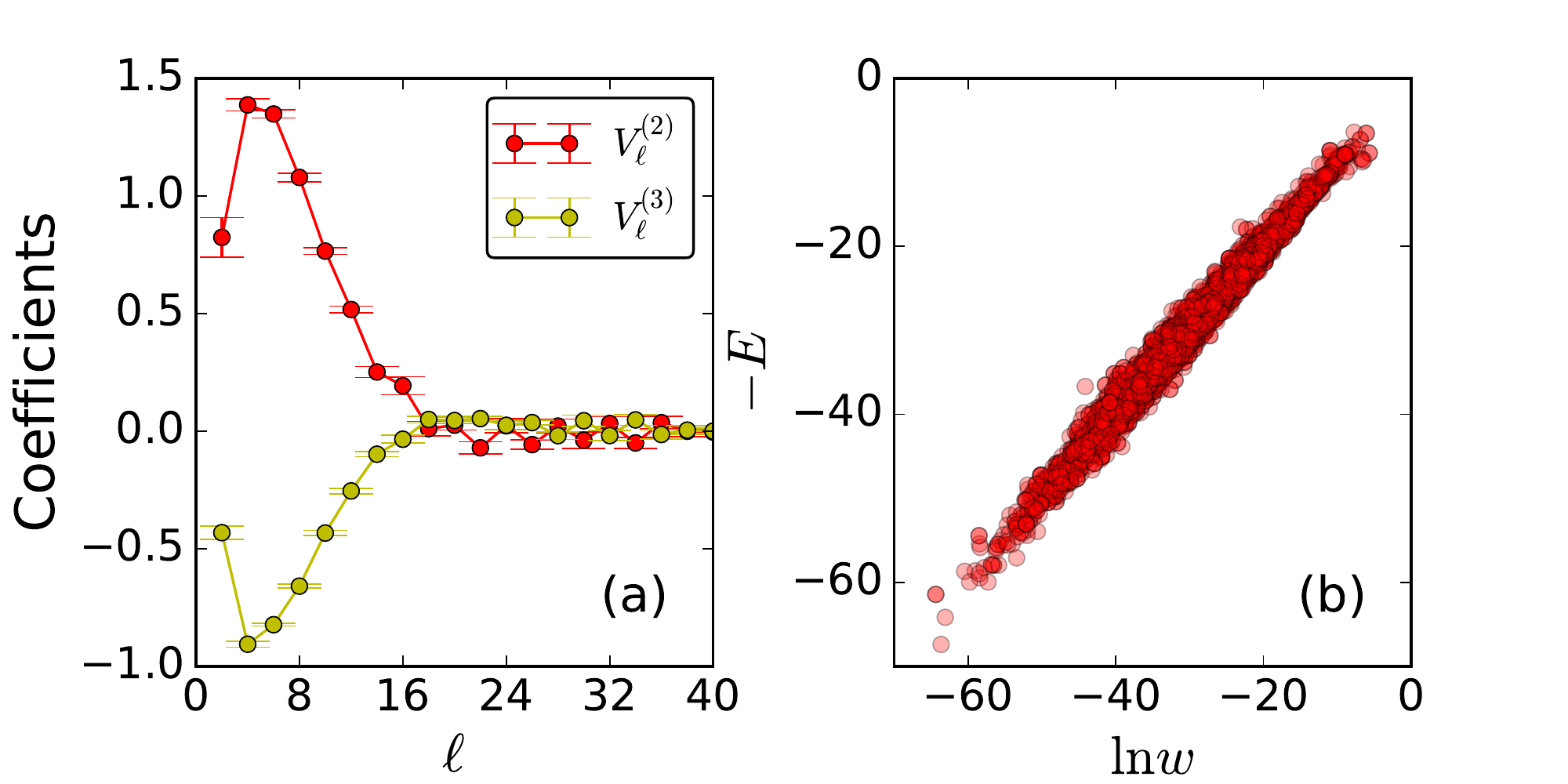}
\caption{(Color online) (a) The Legendre coefficients $\mathcal{V}^{(2,3)}_{\ell}$ of the two and three-body interactions in \Eq{eq:classicalgas}. (b) The log-weight of the CT-QMC~\Eq{eq:expansion} and of the molecular gas model~\Eq{eq:classicalgas}. Each red dot represents a test sample which is not used for training. The physical parameters of the quantum impurity model (\ref{eq:siam}) are $\beta=100, U=-2,\varepsilon=0.2 $ and $ \lambda=1.0$. 
\label{fig:Vlandlogw}}
\end{figure}

The interaction expansion CT-QMC impurity solver~\cite{Rubtsov:2005iw,Gull:2011jd} performs diagrammatic expansion of the partition function in terms of the interaction strength
\begin{equation}
Z/ Z_{0}  =  \sum_{k=0}^{\infty} \int_{0}^{\beta}d\tau_{1} \int_{\tau_{1}}^{\beta}d\tau_{2} \ldots\int_{\tau_{k-1}}^{\beta}d\tau_{k}\, (-U)^{k}|\det\left(G\right)|^{2}, % \nonumber \\ & = & \sum_{k=0}^{\infty}\sum_{\mathcal{C}_{k}}\, w(\mathcal{C}_{k}) ,  
\label{eq:expansion}
\end{equation}
where $Z_{0}$ is the non-interacting partition function. Introducing the configuration $\mathcal{C}=\left\{\tau_{1}, \tau_{2},\ldots, \tau_{k}\right\}$ and the weight $w(\mathcal{C})= (-U)^{k}|\det(G)|^{2}$, we rewrite \Eq{eq:expansion} as $Z/Z_{0} = \sum_{\mathcal{C}}\, w(\mathcal{C})$. Here the summation over configurations denotes the discrete summation over the expansion order and the time-ordered integrations over the imaginary times. $G$ is a $k\times k$ matrix whose matrix elements are given by the non-interacting Green's function $G_{ij}=\mathcal{G}_{0}(\tau_{i}-\tau_{j})-\delta_{ij}/2$. The CT-QMC simulation of the single impurity Anderson model is not hindered by the fermion sign problem in general~\cite{Anonymous:2005ib}. For simplicity, in the following we consider $U<0$ so that we can directly interpreted $w(\mathcal{C})$ as a positive Boltzmann weight~\footnote{The repulsive case can be handled with the trick of Refs.~\cite{Rubtsov:2005iw,Assaad:2007be}. Note that we also use a finite local chemical potential $\varepsilon$ to tune away from the special particle-hole symmetric point, where all the odd order contributions in \Eq{eq:expansion} vanish~\cite{Rubtsov:2005iw}.}. Accelerating the CT-QMC methods with a recommender engine is nevertheless detached from the issue of the sign problem because one can always model the probability distribution $|w(\mathcal{C})|$ and gain speedups. 

The expansion Eq.~(\ref{eq:expansion}) formally maps the zero-dimensional quantum impurity model (\ref{eq:siam}) to a one-dimensional ``classical molecular gas'' model shown in the Fig.~\ref{fig:mapping}(b)~\cite{Wang:2015bva,Wang:2015je}. The molecular gas is in the grand canonical ensemble, where each molecule represents an interaction vertex resides in the continuous imaginary-time axis. Conventional updates of the CT-QMC methods~\cite{Rubtsov:2005iw,Gull:2011jd} indeed resemble the grand canonical Monte Carlo simulation of the molecular gasses~\cite{leach2001molecular, frenkel2002understanding}. In these simulations, one attempts to insert or remove a vertex according to a uniform probability distribution and accepts or rejects the move according to the change of the Monte Carlo weight. However, these simple updates ignore the correlations between the vertices and can suffer from low acceptance rates and long autocorrelation times. Curing such inefficiency requires a better analytical understanding of the correlations in the Monte Carlo configuration and designing suitable updates correspondingly. This is, however, a nontrivial task because of the determinant in the CT-QMC weight~(\ref{eq:expansion}). 
%Moreover, compared to the conventional molecular simulations, the CT-QMC method has an unfavorable $\mathcal{O}(k^{3})$ scaling due to the determinant calculations. 

\begin{figure}[t]
\includegraphics[width=9cm]{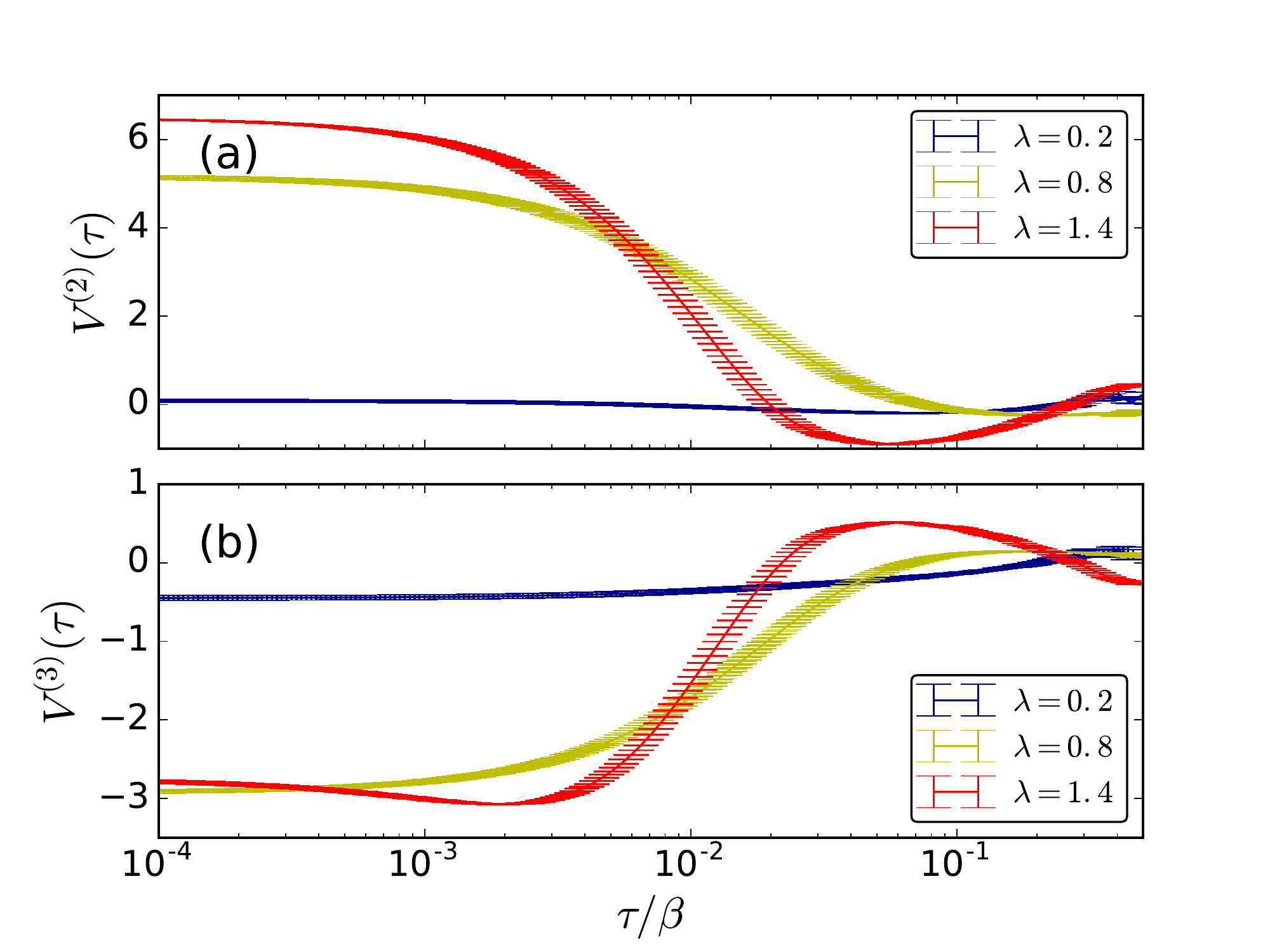}
\caption{(Color online) (a) The two-body and (b) the three-body interaction potentials for various hybridization strengths $\lambda$. Positive (negative) value means attractive (repulsive)  interactions due to the minus sign in \Eq{eq:classicalgas}. The physical parameters are identical to the Fig.~\ref{fig:Vlandlogw}.   
\label{fig:Vtau}}
\end{figure}

To address these problems we exploit the aforementioned intrinsic quantum to classical mapping explicitly. First, we distill the correlations in the QMC configurations into a classical reference system. Then, we use the reference system as a recommender engine to guide future QMC sampling. In line with the mapping of Fig.~\ref{fig:mapping}(b), we write the partition function of the molecular gas as $Z_\mathrm{gas}=\sum_{\mathcal{C}}e^{-E(\mathcal{C})}$ and assume an explicit form for the energy function of the molecular gas

\begin{equation}
E(\mathcal{C}) = - \sum_{i=1}^k \mathcal{V}^{(2)} (\tau_{i+1} - \tau_{i})- \sum_{i=1}^k\mathcal{V}^{(3)} (\tau_{i+1} - \tau_{i-1}) -\mu k -b. 
\label{eq:classicalgas}
\end{equation}
This form of energy potentials respect the translational invariance of the original problem in the imaginary time axis. $\mathcal{V}^{(2)}$ is a two-body interaction potential depending on the time difference of the two adjacent vertices. Here the subscripts of the imaginary times and the time differences all take into account of the periodic boundary condition of the imaginary time shown in Fig.~\ref{fig:mapping}(b). The second term of (\ref{eq:classicalgas}) is a three-body interaction where the two vertices interact with $\mathcal{V}^{(3)}$ only if there is the third vertex in between. The effective chemical potential $\mu$ term controls the average molecule number in the grand canonical ensemble. Finally, $b$ is an energy offset which controls the relative magnitude of $Z/Z_{0}$ and $Z_\mathrm{gas}$. Equation (\ref{eq:classicalgas}) defines an energy based model for the CT-QMC probability distribution. To determine its exact form, we adopt a data driven point of view and train the model parameters using collected CT-QMC configuration data. Note that the model (\ref{eq:classicalgas}) does not need to reproduce the probabilities exactly. Capturing the crucial correlations in the original CT-QMC configurations is already good enough to be a useful recommender engine.

To determine the interaction potentials, we parametrize these continuous functions using the Legendre polynomial basis
$\mathcal{V}^{(2,3)}(\tau)  = \sum_{\ell = 1}^L  P_{\ell} [x (\tau)]\mathcal{V}^{(2,3)}_{\ell}$, where $P_{\ell}[x]$ is the $\ell$-th order Legendre polynomial and $x (\tau) = 2 \tau / \beta - 1$ maps the continuous time differences $\tau$ to the region $[-1, 1]$. We keep the expansion coefficients $\mathcal{V}_{\ell}^{(2,3)}$ up to $L$-th order. Periodicity of the imaginary-time axis implies $\mathcal{V}^{(2,3)}(\tau) = \mathcal{V}^{(2,3)}(\beta -\tau)$. Thus, we keep only even $\ell$ terms in the Legendre expansion~\footnote{Note that we have also exclude the constant term $\ell=0$ in the expansion since they can be absorbed into the chemical potential term.}. Requiring $Z_\mathrm{gas}$ to match the expansion of $Z/Z_{0}$ term by term,  
%$w(\mathcal{C}_{k}) =e^{-E(\mathcal{C}_{k})}$, 
we have  
\begin{eqnarray}
%\ln[w(\mathcal{C}_{k})] =  \sum_{\ell = 1}^L \left[\frac{\sqrt{2 \ell + 1}}{\beta} \sum_{i < j}^k P_{\ell} [x(\tau_i - \tau_j)] \right]\mathcal{U}_{\ell} +  \sum_{\ell = 1}^L \left[\frac{\sqrt{2 \ell + 1}}{\beta} \sum_{i < j}^k P_{\ell} [x(\tau_i - \tau_j)] \right]\mathcal{V}_{\ell}  + \mu k + E_0
\ln[w(\mathcal{C})] & = & \sum_{\ell = 1}^L   \left\{  \sum_{i=1}^k  P_{\ell} [x(\tau_{i+1} - \tau_{i})] \right\}\mathcal{V}^{(2)}_{\ell}  \nonumber\\ & + &  \sum_{\ell = 1}^L \left\{ \sum_{i=1}^k  P_{\ell} [x(\tau_{i+1} - \tau_{i-1})] \right\}\mathcal{V}^{(3)}_{\ell} \nonumber \\ &+& \mu k  + b. 
\label{eq:fit}
\end{eqnarray} 
The equation (\ref{eq:fit}) defines a linear regression problem for the parameters $\left\{\mathcal{V}^{(2)}_{\ell}, \mathcal{V}^{(3)}_{\ell}, \mu, b\right\}$. From a machine learning perspective, the Legendre polynomials $P_{\ell}[x]$ and the expansion order $k$ are the features we manually extracted from the CT-QMC configuration $\mathcal{C}=\left\{\tau_{1}, \tau_{2},\ldots, \tau_{k}\right\}$. This feature engineering is motivated by the physical considerations based on the molecular gas model (\ref{eq:classicalgas}). The quantum to classical mapping naturally solves the problem of modeling a fluctuating number of continuous random variables.  
 
To collect the training data, we perform CT-QMC simulations with conventional random insertion and removal updates~\cite{Rubtsov:2005iw}. For each update whether it is accepted and rejected we extract the features in the right hand of \Eq{eq:fit} and compute the log-weight as the regression target. After collecting around $20,000$ samples we perform the ridge regression~\cite{esl2009} for the fitting parameters, where we use a $L_{2}$ regularization of the strength $10^{-3}$ for the coefficients $\left\{\mathcal{V}^{(2)}_{\ell}, \mathcal{V}^{(3)}_{\ell}, \mu \right\}$ to prevent overfitting. Figure~\ref{fig:Vlandlogw}(a) shows the fitted Legendre coefficients. The error bars are estimated using eight independent runs. The coefficients vanish for large $\ell$, justifying the truncation of the Legendre expansion~\footnote{Higher order Legendre polynomials fit the detailed oscillatory behavior of the interaction potential and potentially lead to overfitting.}. To verify the fitting, Fig.~\ref{fig:Vlandlogw}(b) shows the exact log-weight of the CT-QMC and the predicted log-weight of the classical gas model (\ref{eq:classicalgas}) on the test samples. Strong positive correlation indicates that the fitting indeed captures the distribution of the CT-QMC configurations well. 

Figure~\ref{fig:Vtau} shows the effective interaction potentials of the  molecular gas model (\ref{eq:classicalgas}). For weak hybridization strength, the molecules are effectively non-interacting because there is very little correlations in the imaginary time for such a nearly isolated quantum impurity. As the hybridization strength $\lambda$ increases, the two-body interaction potential $\mathcal{V}^{(2)}$ becomes attractive while the three-body interaction $\mathcal{V}^{(3)}$ becomes repulsive near short time differences. The combined interaction effects will favor configurations with bounded pairs of vertices. %Including the three-body interaction in \Eq{eq:classicalgas} is crucial to model the CT-QMC distribution because it prevents the molecules to collapse together. 
Physically, various impurity quantum phase transitions and crossovers manifest themselves in the classical gas model~\cite{Wang:2015bva, Wang:2015je, PhysRevB.94.235110}. Computationally, knowing the effective interactions between the vertices can help us sample them more efficiently.

\begin{figure}[t]
\includegraphics[width=\columnwidth]{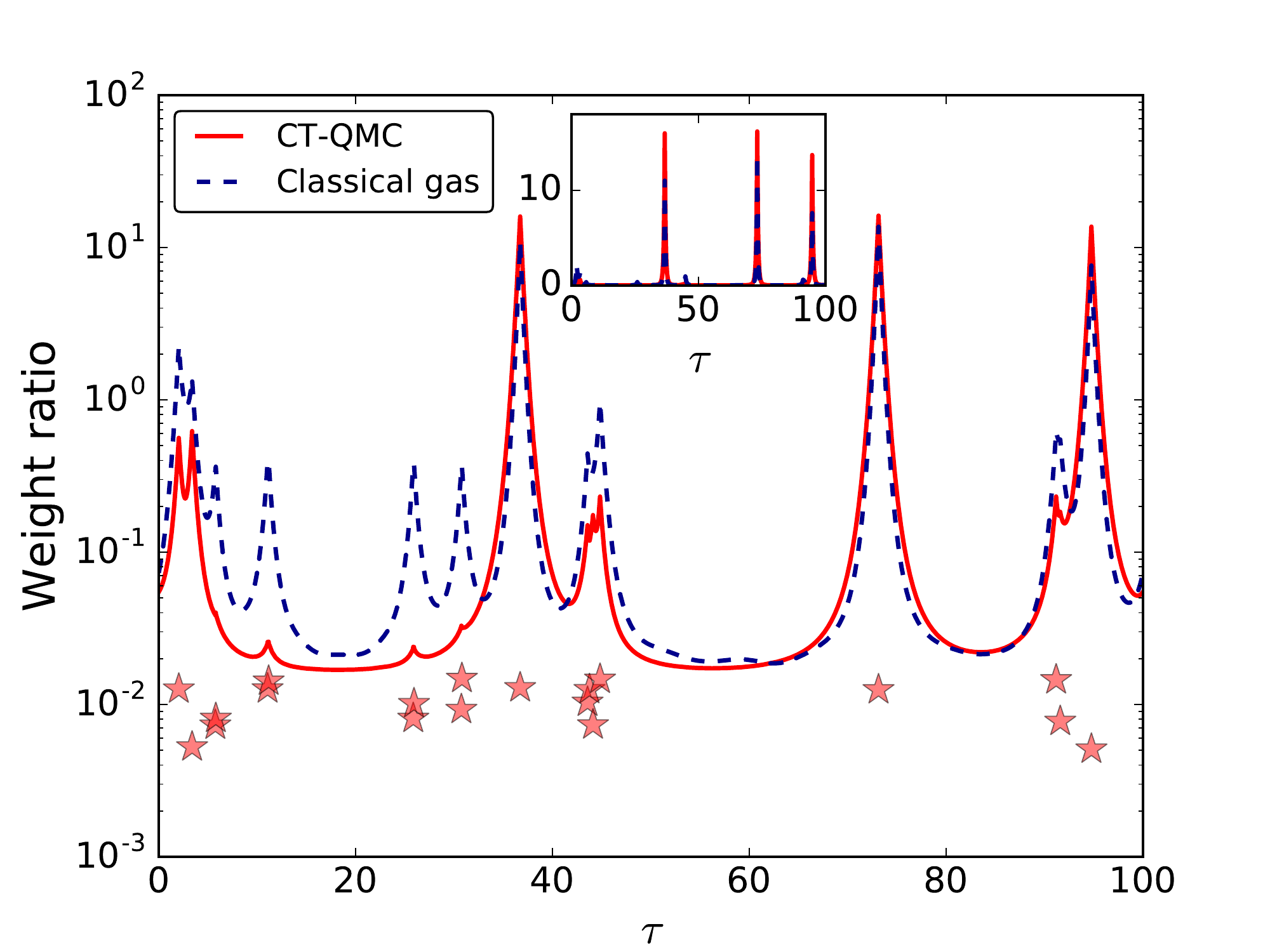}
\caption{(Color online) The weight ratio of adding a vertex at the imaginary time $\tau$ to the configuration $\mathcal{C}$. The red solid line is $w(\mathcal{C}\cup\{\tau\})/w(\mathcal{C})$, the blue dashed line is $e^{-E(\mathcal{C}\cup\{\tau\})+E(\mathcal{C})}$. The red stars indicate the location of the existing vertices in $\mathcal{C}$. For visibility we offset their vertical positions in the graph. Inset shows the same plot in the linear scale which highlights the three dominant peaks. The physical parameters $\beta,U,\varepsilon$ are identical to the Fig.~\ref{fig:Vlandlogw}.  
\label{fig:accratio}}
\end{figure}
 
To better understand how well the classical molecular gas model captures the correlation between the interaction vertices, Fig.~\ref{fig:accratio} shows the weight ratio of adding a vertex at the imaginary-time $\tau$. This ratio determines the acceptance rate of the insertion update. The red stars indicate the locations of the existing vertices $\mathcal{C}$. The weight ratio exhibits peaks around them; some peaks are more pronounced than the others. This is because of the effective two-body attractive interactions enhances the probability of adding the vertex near the existing vertices. While the effective three-body repulsive interaction suppresses the probability of adding the third vertex in the vicinity of two already paired-up vertices. For example, the CT-QMC weight ratio only exhibits a small peak around $\tau\sim 11$ in Fig.~\ref{fig:accratio}. The classical model \Eq{eq:classicalgas} correctly capture this crucial three-body correlation of the interaction vertices. Without the three-body interaction term, the vertices will collapse into clusters which are certainly not favored in the original CT-QMC simulation. The overall effect of the combined interactions is that there are only three dominant peaks in the weight ratio around the three isolated vertices, see the inset of Fig.~\ref{fig:accratio}. Randomly inserting a vertex in $[0, \beta)$ without taking into account of this highly nonuniform distribution will have poor acceptance probability. 

%Since the trained classical gas model (\ref{eq:classicalgas}) models the true distribution well, we use it the to speed up the CT-QMC simulation. 
Having trained the classical gas model as a proxy of the original probability distribution, we use it as a recommender system for the QMC simulation. In general, we simulate the classical molecular gas using efficient Monte Carlo methods and recommend the update back to the original QMC simulation. Assuming the simulation of the classical gas model satisfies the detailed balanced condition, we accept the recommended move from the configuration $\mathcal{C}$ to $\mathcal{C}^{\prime}$ with the acceptance probability~\cite{Huang:2016tf, 1611.09364},  
\begin{equation}
A(\mathcal{C}\rightarrow \mathcal{C}^{\prime}) = \min\left\{1,\frac{e^{-E(\mathcal{C})}}{e^{-E(\mathcal{C}^{\prime})}}\cdot\frac{w(\mathcal{C}^{\prime})}{w(\mathcal{C})}\right\}.
\label{eq:accratio}
\end{equation}
Equation (\ref{eq:accratio}) guarantees an unbiased simulation with improved acceptance ratio. This approach boosts the overall performance because of proposing more probable updates. 

There are various ways that the CT-QMC simulation can benefit from the recommender system~\cite{Huang:2016tf, 1610.03137, 1611.09364}. First, the updates can be nonlocal, in the sense that $\mathcal{C}^{\prime}$ can differ drastically from $\mathcal{C}$ while still keeping a high acceptance rate given an accurate fitting in \Eq{eq:fit}. Furthermore, even without the luxury of performing global updates for the reference system, one can still afford to accumulate many local update steps before recommending a nonlocal update to the CT-QMC simulation. Because the simulation of the molecular gas is much cheaper than the CT-QMC ($\mathcal{O}(1)$ versus $\mathcal{O}(k^{2})$ operations per local update step~\footnote{After the fitting \Eq{eq:fit} is done we precompute the interaction potentials $\mathcal{V}^{(2,3)}(\tau)$ on a fine mesh and use linear interpolations to obtain their value at other imaginary times during the molecular simulation.}), the recommendation step has little overhead. Finally, as long as the classical molecular gas model captures certain correlations in the CT-QMC configurations, it would already be beneficial to exploit this information and design better update proposals. In the last case, the recommended update can still be local, but has  an improved acceptance rate and enjoys the advantage of the $\mathcal{O}(k^{2})$  fast update scheme in the CT-QMC~\cite{Rubtsov:2005iw, Gull:2011jd}. 

\begin{figure}[t]
\includegraphics[width=\columnwidth]{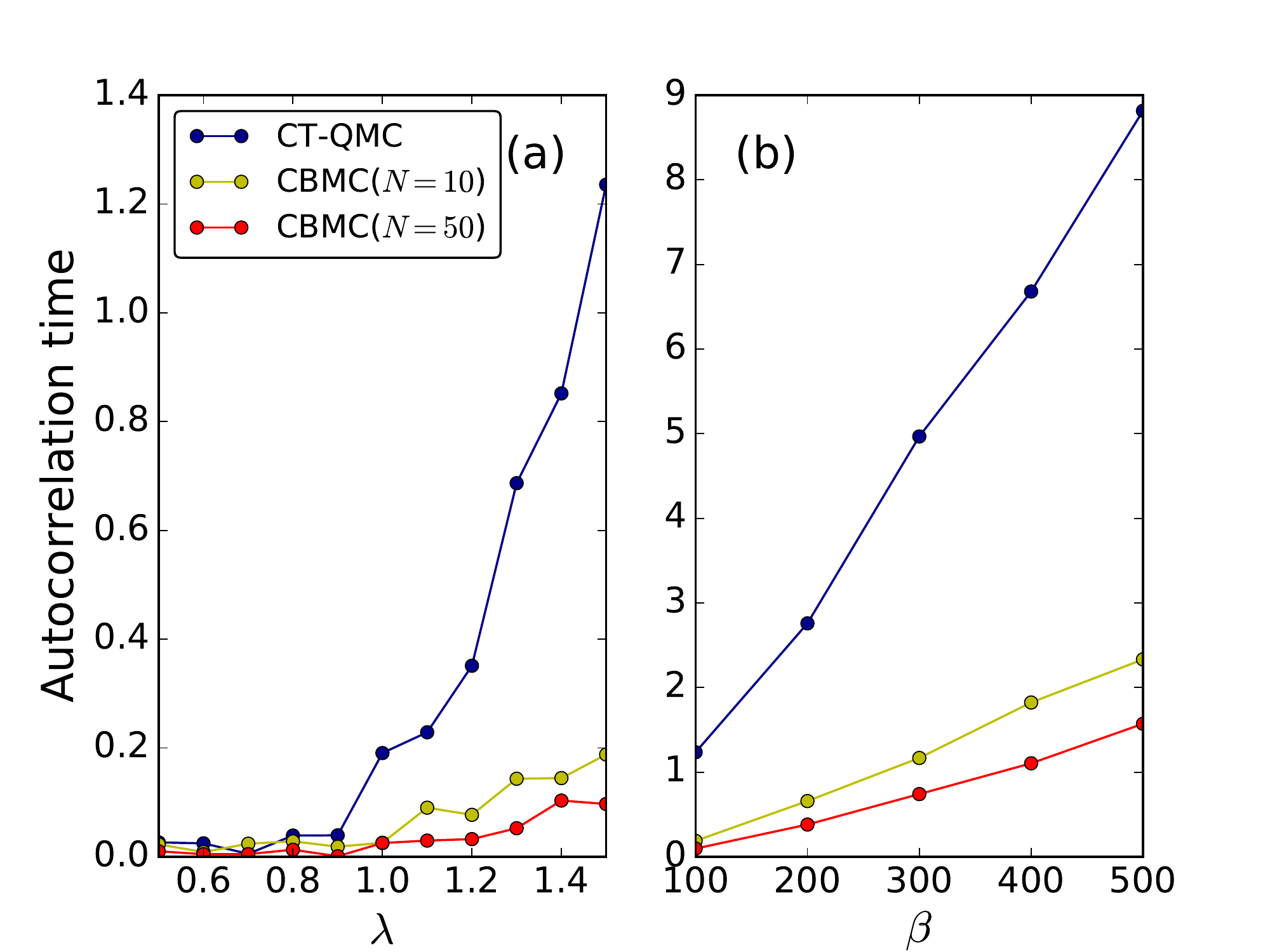}
\caption{(Color online) The autocorrelation times of the total expansion order. The improvement of the ordinary CT-QMC results (blue dots)~\cite{Rubtsov:2005iw, Gull:2011jd} is more significant in the difficult parameter region where the hybridization strength $\lambda$ or the inverse temperature $\beta$ is large. Increase the trail steps $N$ in the CBMC simulation of the molecular gas model (\ref{eq:classicalgas}) further reduces the autocorrelation time. In (a) $\beta=100$ and in (b) $\lambda=1.5$, the other physical parameters $U,\varepsilon$ are identical to the Fig.~\ref{fig:Vlandlogw}.  
\label{fig:improve}}
\end{figure}

We employ the configurational bias Monte Carlo (CBMC) method~\cite{Siepmann:1992kf} to simulate the molecular gas model (\ref{eq:classicalgas}). CBMC is an efficient molecular simulation technique~\cite{leach2001molecular, frenkel2002understanding}, which is particularly useful for growing long molecular chains. The basic idea of CBMC is to probe the landscape around the current configuration and find a move with higher acceptance rate. To achieve this goal, we perform $N$ independent trial updates from the old configuration and select an actual one according to their relative probabilities~\cite{SM}. We then propose this CBMC update to the CT-QMC and accept it with the probability (\ref{eq:accratio}). The number of trail steps $N$ in the CBMC controls how much information we'd like to extract from the molecular gas recommender system.  

As a relevant measure of the improvement in the Monte Carlo sampling, Fig.~\ref{fig:improve} shows the autocorrelation times of the expansion order measured in the unit of $10^{3}$ CT-QMC update steps. Taking the recommendations from the CBMC simulation greatly reduces the autocorrelation time, especially in the challenging parameter regions with strong hybridization strength and at low temperature. It is encouraging to see that  a few trial steps ($N=10$) in the CBMC already significantly improves the efficiency of the CT-QMC. Moreover, one can afford even a larger number of CBMC trail steps because computing the weight ratio of \Eq{eq:classicalgas} is much cheaper than \Eq{eq:expansion}. Increasing $N$ further improves the autocorrelation time by finding more probable updates which respect the correlations of the CT-QMC configurations. 

Since the expansion order is a global property of the CT-QMC configuration, its autocorrelation time is typically larger than the other physical observables. Greatly reducing this global autocorrelation time in the CT-QMC gives the hope of computing the fidelity susceptibility of correlated fermions at substantially lower temperature and larger system sizes~\cite{Wang:2015bva, Wang:2015je, PhysRevB.94.235110}. 

%The acceptance rate even doubles in the most challenging parameter region with strong hybridizations. 

%Now, since we have approximated the determinant with the more tracble expression.  
%Without the fitted classical model, such trails are unaffordable because computing the determinant ratio costs $\mathcal{O}(k^{2})$.

%outlook
Explicitly constructing the effective classical model and use it as a recommender system is a general approach to speed up the QMC simulations. Identifying the classical molecular gasses as the recommender engines for CT-QMC methods brings a large number of powerful molecular simulation techniques~\cite{leach2001molecular, frenkel2002understanding} into the game. In this paper we employ the CBMC method~\cite{Siepmann:1992kf} to efficiently explore the probability landscape of the classical molecular gas model. One may consider using other successful molecular simulation approaches such as the hybrid Monte Carlo~\cite{Duane:1987uq} or geometric cluster algorithms~\cite{Liu:2004eb} for further improvements.

For other variants of the CT-QMC methods with auxiliary fields in the configurations~\cite{Rubtsov:2005iw, Assaad:2007be, Gull:2008cm}, it is natural to generalize the effective interaction potentials in \Eq{eq:classicalgas} to be dependent on the auxiliary fields degree of freedoms in addition to the imaginary times. The present recommender engine strategy can also be generalized to the hybridization expansion~\cite{Werner:2006ko} and the Kondo coupling expansion~\cite{Otsuki:2007ff} CT-QMC algorithms. In the later case, the spin-flip events in the imaginary time map to the classical Coulomb charges~\cite{Wang:2015je} according to the seminal work of Anderson and Yuval~\cite{Anderson:1969wl}. Going beyond the quantum impurity models, the "recommender engine" approach can benefit a broad range of modern QMC methods for interacting bosons~\cite{PhysRevLett.77.5130, Prokofev:1998tc, Evertz:2003ch,Kawashima:2004clb}, quantum spins~\cite{Sandvik:1991tn,1999PhRvB..5914157S}, and lattice fermions~\cite{Iazzi:2015hi, 2015PhRvB..91w5151W,  Liu:2015kx}. In those cases, the classical molecules will carry additional indices to indicate the spatial location of the interaction events.

Besides serving as a recommender engine to speed up the QMC simulations, the classical reference system does capture physical information of the original quantum problem. For example, the average particle number is related to the average interaction energy; while the classical compressibility is related to the second order derivative of the quantum free energy; and the bipartite particle fluctuation~\cite{Rachel:2012eu} of the classical system corresponds to the fidelity susceptibility of the quantum system~\cite{Wang:2015bva}. These discussions indicate that the phase transitions of the quantum system will always manifest themselves in the corresponding classical representations. The correspondence calls for special attentions in designing the classical reference system. For example, it is known that a one dimension model with the nearest neighbor interaction only has no phase transition~\cite{takahashi1942simple}. Therefore, to ensure the classical reference system has enough descriptive power one may need allow additional internal degree of freedoms or longer ranged interactions between the molecules. 
 
%Sign problem
The recommender engine approach works as well even if the sign of the weight is not positive definite. In this case, sampling according to the absolute value of the weight still defines a legitimate statistical mechanics problem. However, the physics of the classical reference model may be detached from the original quantum system~\cite{Anonymous:PxlD92x3,Anonymous:jxjVcLI0}. 

\begin{acknowledgments}
L.W. is supported by the Ministry of Science and Technology of China under the Grant No.2016YFA0302400 and the start-up grant of IOP-CAS. L.H. is supported by the Natural Science Foundation of China No.11504340 and the Foundation of President of China Academy of Engineering Physics (No.~YZ2015012). Y.Y. is supported by the National Natural Science Foundation of China (No. 11522435) and the Strategic Priority Research Program (B) of the Chinese Academy of Sciences (No. XDB07020200). We use the ALPS library~\cite{BBauer:2011tz} for the Monte Carlo data analysis. We thank Zi Cai for proof reading of the manuscript. 
\end{acknowledgments} 

\bibliographystyle{apsrev4-1}
\bibliography{ML-CTQMCpaper}

\clearpage

%\section{Linear regression with $L_{2}$-regularization}
%For the convenience of the readers we provide the formula for the linear regression fit. The details can be found in the textbooks on machine learning~\cite{..}. 
%
%We minimize the target function 
%\begin{equation}
%\chi^{2}= \sum_{i=1}^{N} \left(y_{i}-b-\sum_{j=1}^{p}x_{ij} w_{j}\right)^{2} + \eta \sum_{j=1}^{p}w_{j}^{2}
%\end{equation}
%where we use a $L_{2}$-regularization parameter $\eta=0.001$. 
%
%The solution for the fitting parameters is obtained by perform the singular value decomposition of the data matrix $X = UDV^{T}$, 
%
%\begin{equation} 
%w_{j}=V \frac{s}{s^{2}+\eta} U^{T}y
%\end{equation}

\section{Supplementary Materials}

\subsection{Configurational bias Monte Carlo simulation for the molecular gas model}
We present details of the configurational bias Monte Carlo~\cite{Siepmann:1992kf} simulation of the classical molecular gas model $Z_\mathrm{gas}=\sum_{\mathcal{C}}e^{-E(\mathcal{C})}$. Pedagogical introductions about the CBMC method can be found in the textbooks of the molecular simulations~\cite{leach2001molecular, frenkel2002understanding}.
 
For the insertion update, we randomly generate $N$ imaginary times in the range $\tau_{i}\in[0, \beta)$ and compute the corresponding Boltzmann weight ratios $r_{i} = e^{-E(\mathcal{C}\cup\{\tau_{i}\})+E(\mathcal{C})}$. We then select a $\tau_{i}$ according to the discrete distribution ${r_{i}}/{W}$, where $W=\sum_{i=1}^{N} r_{i}$. Assuming the current configuration $\mathcal{C}$ contains $k$ imaginary times, the acceptance rate of the insertion update is 

\begin{equation}
A\left(\mathcal{C}\rightarrow \mathcal{C}\cup\{\tau_{i}\}\right) = \min\left\{1, \frac{\beta W}{(k+1) N}  \right\}.
\end{equation}

For the removal update, we randomly select a vertex from the $k$ existing vertices in the current configuration $\mathcal{C}$. Suppose its imaginary-time is $\tau$, we compute $W=e^{-E(\mathcal{C})+ E(\mathcal{C}\setminus\{\tau\})} +\sum_{i=2}^{N} e^{- E(\mathcal{C}\setminus\{\tau\}\cup\{\tau_{i}\})+E(\mathcal{C}\setminus\{\tau\}) }$ with $N-1$ random numbers draw from $\tau_{i}\in[0, \beta)$. The acceptance rate reads 

\begin{equation}
A\left(\mathcal{C}\rightarrow \mathcal{C}\setminus\{\tau\}\right)  = \min\left\{1, \frac{k N}{\beta W}  \right\}. 
\end{equation}
In the case of $N=1$, the above algorithm reduce to the ordinary single particle insertion/removal update of a molecular gas in the a grand canonical ensemble. With increasing $N$, one will find more probable update according to the probability distribution $e^{-E(\mathcal{C})}$. Finally, we recommend the configuration from the CBMC update to the CT-QMC simulation. Notice that the update remains local for any choice of $N$. 

\end{document}